\pdfoutput=1
\documentclass[reprint,english,twocolumn,prl,showpacs,superscriptaddress,noeprint,preprintnumbers,nofootinbib,floatfix]{revtex4-1}
\usepackage[T1]{fontenc}
\usepackage[latin1]{inputenc}
\usepackage{graphicx}
\usepackage{amssymb}
\usepackage{color}
\usepackage{units}
\usepackage{babel}
\usepackage{longtable}
\usepackage{amsmath}
\usepackage{subfigure}
\usepackage{paralist}

\newcommand{\p}[1]{\ensuremath{#1^{\prime}}}

\newcommand{\bitem}{\begin{itemize}}
\newcommand{\eitem}{\end{itemize}}
\newcommand{\benum}{\begin{enumerate}}
\newcommand{\eenum}{\end{enumerate}}

 \newcommand{\beqn}{\begin{equation}}
 \newcommand{\eeqn}{\end{equation}}

\def\bea{\begin{eqnarray}}
\def\eea{\end{eqnarray}}

\definecolor{Green}{rgb}{0.1,0.8,0.1}

\definecolor{red}{RGB}{200,20,20}
\definecolor{green}{RGB}{20,200,20}
\definecolor{blue}{RGB}{20,20,200}

\begin{document}

\preprint{
\vbox{
\null \vspace{-0.2in}
\hbox{LTH 997} 
\hbox{MS-TP-14-01} 
\hbox{LPSC 14-007}
}
}

\title{Can new heavy gauge bosons be observed
in ultra-high energy cosmic neutrino events?}

\author{T. Je\v{z}o}
\affiliation{Department of Mathematical Sciences, University of Liverpool, 
	Liverpool L69 3BX, UK}

\author{M. Klasen}
\affiliation{Institut f\"ur Theoretische Physik, Universit\"at M\"unster,
D-48149 M\"unster, Germany
}

\author{F. Lyonnet}
\affiliation{LPSC, Universit\'e Joseph Fourier/CNRS-IN2P3/INPG,
  UMR5821, F-38026 Grenoble, France}

\author{F. Montanet}
\affiliation{LPSC, Universit\'e Joseph Fourier/CNRS-IN2P3/INPG,
  UMR5821, F-38026 Grenoble, France}

\author{I. Schienbein}
\affiliation{LPSC, Universit\'e Joseph Fourier/CNRS-IN2P3/INPG,
  UMR5821, F-38026 Grenoble, France}

\author{M. Tartare}
\affiliation{LPSC, Universit\'e Joseph Fourier/CNRS-IN2P3/INPG,
  UMR5821, F-38026 Grenoble, France}


\begin{abstract}
A wide range of models beyond the Standard Model predict
 charged and neutral resonances, generically called $W'$- and $Z'$-bosons, respectively. 
In this paper we study the impact of such resonances 
on the deep inelastic scattering of ultra-high energy neutrinos as well as on the resonant charged current $\bar\nu_e e^-$ 
scattering (Glashow
resonance). 
We find that  the effects of such resonances can not be observed with the Pierre Auger Observatory or any foreseeable upgrade of it.
\end{abstract}


\maketitle


New charged and neutral resonances are predicted in many well-motivated extensions of the
Standard Model (SM) such as theories of grand unification (GUTs) or models
with extra spatial dimensions \cite{Langacker:2008yv}.
These extensions generally do not predict the precise energy scale at which the new heavy
states should manifest themselves. However, for various theoretical reasons
(e.g.\ the hierarchy problem) new physics is expected to appear at the TeV scale and is
searched for at the Large Hadron Collider (LHC) which will soon operate at a center-of-mass
energy of $\sqrt{s}= 13$ TeV.
At the same time, important restrictions on new physics scenarios are imposed by low-energy
precision observables.
On the other hand, highly energetic interactions of cosmic rays in the atmosphere involve processes
at higher center-of-mass energies than those reached by the LHC.
Motivated by this fact, we study the prospects to observe new spin-1 resonances in collisions
of ultra-high energy (UHE) neutrinos with nuclei in the atmosphere as analyzed by the
Pierre Auger Collaboration or a future neutrino telescope.
For example, for neutrinos with an energy of about 10$^{19}$ eV, 
the center-of-mass energy of the neutrino-nucleon interactions is about $\sqrt{s} \simeq 140$ TeV,
considerably extending the energy range accessible at the LHC.
So far, no UHE neutrino events have been observed by the Pierre Auger Observatory
which has led to improved limits on the diffuse flux of UHE neutrinos
in the energy range $E_\nu \ge 10^{18}$ eV \cite{Abreu:2013ppa,Abreu:2012zz}.

The potential of the Pierre Auger Observatory for testing new physics scenarios like extra dimensions or the formation 
of micro-black holes has been studied in \cite{Giesel:2003hj} and \cite{Anchordoqui:2005ey}. 
In this report we revisit the predictions for cross sections 
in the SM, and we explore the
impact of new charged ($W'$) and neutral ($Z'$) gauge bosons 
on these quantities. We address the following questions:
\begin{inparaenum}[(i)]
\item
Assuming the LHC does observe new charged or neutral spin-1 resonances,
how would this affect the predicted neutrino cross sections?
\item
Assuming the LHC does not discover any new spin-1 resonances,
what are the prospects to observe heavy $W'$- and $Z'$-bosons with masses larger
than 5 TeV using UHE cosmic neutrino events?
\end{inparaenum}

For definiteness, we consider $W'$ and $Z'$ bosons due to an extended 
$\text{G}(221) \equiv \text{SU}(2)_{1}\times \text{SU}(2)_{2}\times \text{U}(1)_{X}$ gauge group.
In this framework, constraints on the parameter space from low-energy
precision observables have been derived in \cite{Hsieh:2010zr} and the
collider phenomenology has been studied in \cite{Jezo:2012rm,Cao:2012ng,Abe:2012fb,Jinaru:2013eya}.
Several well-known models emerge naturally from different ways of
breaking the $\text{G}(221)$ symmetry down to the SM gauge group
\cite{Hsieh:2010zr}, in particular
Left-Right (LR) \cite{Mohapatra:1974gc,Mohapatra:1974hk,Mohapatra:1980yp},
Un-Unified (UU) \cite{Georgi:1989ic,Georgi:1989xz}, 
Non Universal (NU) \cite{Malkawi:1996fs,Li:1981nk}, 
Lepto-Phobic (LP), Hadro-Phobic (HP) and Fermio-Phobic (FP) \cite{Barger:1980ix,Barger:1980ti}
models.
In addition, we present results for the Sequential Standard Model
(SSM) \cite{Altarelli:1989ff}, where the $W'$- and $Z'$-bosons are just
heavy copies of the $W$- and $Z$-bosons
in the SM.
This is motivated by the fact that the SSM often serves as a benchmark model in the 
literature 
\cite{CMS:2013qca,ATLAS:2013jma,CMS:2013rca,Aad:2012dm}.

In the SM, the following neutrino interactions can take place in the atmosphere \cite{Gandhi:1995tf,Gandhi:1998ri}:
\begin{inparaenum}[(i)]
\item Charged current deep-inelastic scattering (CC DIS): 
$\nu_\ell + N \to \ell^- + X$, $\bar\nu_\ell + N \to \ell^+ + X$.
Here, $\nu_\ell$ stands for the three neutrino flavors $\nu_e, \nu_\mu, \nu_\tau$.
\item Neutral current deep-inelastic scattering (NC DIS): 
$\nu_\ell + N \to \nu_\ell + X$, $\bar\nu_\ell + N \to \bar \nu_\ell + X$.
\item The Glashow resonance (GR) \cite{Glashow:1960zz,Berezinsky:1977sf,Berezinsky:1981bt}: 
$\bar\nu_e + e^- \to \bar\nu_\ell + \ell^{-}$, $\bar\nu_e + e^- \to q + \bar q'$,
where $q=u,d,s,c,b$.
Obviously, charged current resonant $s$-channel scattering occurs only for incoming 
anti-electron neutrinos. The process $\bar\nu_e + e^- \to \bar \nu_e + e^-$ also
has a non-resonant neutral current $t$-channel contribution.
\item Non-resonant neutrino-electron scattering: 
\begin{inparaenum}
\item 
$\nu_e + e^- \to \nu_e + e^-$, which has contributions from $W$ and $Z$ exchange diagrams.
\item 
Charged current $\nu_\mu e^-$ and $\nu_\tau e^-$ scattering in the atmosphere:
$\nu_\ell + e^-  \to \ell^- + \nu_e$ ($\ell = \mu, \tau$).
Note that the corresponding process with incoming anti-neutrinos is not possible.
\item 
Neutral current scattering of $\nu_\mu$, $\bar\nu_\mu$, $\nu_\tau$, $\bar\nu_\tau$ and $\bar\nu_e$:
$\nu_\ell + e^- \to \nu_\ell + e^-$, $\bar\nu_\ell + e^- \to \bar\nu_\ell + e^-$.
\end{inparaenum}
\end{inparaenum}

In the following, we mainly focus on the dominant cross sections of neutrino--nucleon DIS 
and neglect the contributions from non-resonant neutrino--electron scattering which are smaller
by several orders of magnitude. 
The $W'$ and $Z'$ resonances contribute to the $\nu N$ DIS,
where the main contribution comes from the interference with the SM amplitudes.
We also consider the Glashow resonance, which has attracted a lot of interest
in the literature as a way to detect extra-galactic neutrinos and as a discriminator
of the neutrino production mechanism/of the relative abundance of the $pp$ and $p\gamma$ sources 
\cite{Anchordoqui:2004eb,Bhattacharjee:2005nh,Hummer:2010ai,Mehta:2011qb,Xing:2011zm,Bhattacharya:2011qu}.
While the GR is entirely negligible at energies $E_\nu \ge 10^8$ GeV there is a new, potentially interesting, resonance 
due to the $W'$-boson which we call GR${}^{\prime}$ in the following.

The differential cross section for DIS mediated by interfering gauge bosons $B$ and $\p{B}$ can be written as 
\begin{equation}
\frac{d^{2}\sigma}{dxdy}=\sum_{B,\p{B}}
\frac{d^{2}\sigma^{B\p{B}}}{dxdy}\, ,
\label{eq:diffContributions}
\end{equation}
where the Bjorken variable $x$ and the inelasticity $y$ are defined as usual.
Furthermore, $B, \p{B} \in \{W, \p{W}\}$ in the case of CC DIS and $B, \p{B} \in \{\gamma, Z, \p{Z}\}$ 
in the case of NC DIS.
Each of these terms can be calculated from the general expression \cite{Aivazis:1994pi}
\begin{align}
\frac{d^{2}\sigma^{B\p{B}}}{dxdy} &=
\frac{2M_pE_\nu G_{B}G_{\p{B}}}{\pi}
\bigg\{ g_{+l}^{B\p{B}} \Big[ xF_1y^{2}+ F_2\left( 1-y\right) \Big] 
\nonumber\\
&\phantom{=} \pm g_{-l}^{B\p{B}}\left[ xF_3y\left(1-\frac{y}{2}\right) \right] \bigg\}\, ,
\label{eq:diffCrossFi}
\end{align}
where the $\pm$ refers to $\nu p$ and $\bar \nu p$ DIS, respectively.
Here, $g_{\pm f}^{B\p{B}}=C^{B}_{f,L}C^{\p{B}}_{f,L}\pm C^{B}_{f,R}C^{\p{B}}_{f,R}$
are (anti-)symmetric combinations of the left- and right-handed gauge boson couplings to the fermions \cite{Jezo:2012rm}
and
$G_B=g_B^{2}/(Q^{2}+M_B^{2})$ is taking into account the
propagator of the gauge boson $B$ with mass $M_B$,  and 
$g_B = \frac{g_{W}}{2\sqrt{2}}$ 
for charged-current interactions and
$g_B = \frac{g_{W}}{2\cos \theta_{W}}$ for neutral-current interactions.
Furthermore, $F_{1,2,3}(x,Q^2)$ are the CC or NC DIS structure functions
which are generally given as convolutions of parton distribution functions
with Wilson coefficients. Here we use the expression in the ACOT scheme
\cite{Aivazis:1994pi,Aivazis:1993kh,Kretzer:1998ju,Stavreva:2012bs} 
neglecting all the quark masses with exception of the top quark mass. 
The latter appears in the bottom quark initiated
contribution to the charged current structure functions in form of a slow rescaling 
prescription where $F_i(x,Q^2) \propto b(\chi,Q^2)+\bar{b}(\chi,Q^2)$ with
$\chi=\frac{Q^{2}+m_{t}^{2}}{2p\cdot q}=x\left( 1+\frac{m_{t}^{2}}{Q^{2}} \right)$.

We now turn to the Glashow resonance, i.e., the contribution
to the cross section from the process $\bar{\nu}_e(p_a) + e^-(p_b) \to f_i(p_1) + \bar{f}_j(p_2)$
mediated by a resonant $W$- or $W'$-boson in the $s$-channel.
The differential cross section can be written as 
\bea
d\sigma^{B\p{B}}&=&
d\Omega \, \times \,
\mathcal{D}\frac{g_{B}^{2}g_{\p{B}}^{2}}{32\pi^{2}s}\, \times \,
\nonumber\\
&& 
\left[(p_{a}\cdot p_{2})(p_{b}\cdot p_{1})(g_{+l}^{B\p{B}}g_{+f}^{B\p{B}}+g_{-l}^{B\p{B}}g_{-f}^{B\p{B}})\right.
\\
&+&
\left. (p_{a}\cdot p_{1})(p_{b}\cdot p_{2})(g_{+l}^{B\p{B}}g_{+f}^{B\p{B}}-g_{-l}^{B\p{B}}g_{-f}^{B\p{B}}) \right] ,
\nonumber
\label{eq:GRsigma}
\eea
where $d\Omega$ is the solid angle of the final state fermion $f_i$ which can be either a quark or a lepton, 
and
\begin{equation}
	\mathcal{D}=\displaystyle{\frac{(s-M_{B}^{2})(s-M_{\p{B}}^{2}) + M_{B}M_{\p{B}}\Gamma_{B}\Gamma_{\p{B}}}{\left[(s-M_{B}^{2})^{2}+M_{B}^{2}\Gamma_{B}^{2}\right]\left[(s-M_{\p{B}}^{2})^{2}+M_{\p{B}}^{2}\Gamma^{2}_{\p{B}}\right]}}\ .
\label{eq:D}
\end{equation}
Here, $\Gamma_{B}$ is the total decay width of a $B$-boson, 
which we approximate
by the sum of its partial decay widths into two fermions\footnote{We estimated using Pythia that the $W'$ decay into a pair of gauge bosons is at the level of 1-2\%. Note that there are regions of parameter space where the decay of the new gauge boson into additional scalars may be significant. However, even in that case this would not affect our conclusions.}
\begin{equation}
	\Gamma_{B} = \sum_{\{f_{i}, f_{j}\}}\Gamma_{B\rightarrow f_{i}\bar{f}_{j}}=\frac{g_{B}^{2}M_{B}g_{+f}^{BB}(f_{i}, \bar{f}_{j})}{6\pi}\ .
	\label{eq:partialwidth}
\end{equation}

Integrating over the solid angle $d\Omega$ and summing over the gauge bosons $B, \p{B} \in \{W, \p{W}\}$ one obtains the total GR cross section
\begin{equation}
	\sigma(s)=\sum_{B,\p{B}}\frac{s}{12\pi}g_{B}^{2}g_{\p{B}}^{2}g_{+l}^{B\p{B}}g_{+f}^{B\p{B}}\mathcal{D}\ .
\end{equation}

We are now in a position to discuss numerical results for the cross sections of 
UHE neutrino interactions in the atmosphere.
For the CC and NC DIS, we consider an isoscalar target and neglect nuclear effects
so that the structure functions are given by the average of the proton and the neutron 
structure functions, $F_{i}=(F^{n}_{i}+F^{p}_{i})/2$. 
As is well-known, the UHE neutrino cross sections in DIS are sensitive to the PDFs at very small momentum fractions $x$
down to $x \simeq 10^{-12}$ which results in large uncertainties as shown in Sarkar et al.~\cite{CooperSarkar:2011pa}.
On the other hand, the UHE neutrino cross sections are quite insensitive to the lower bound for the $Q^2$ integration for which we 
take $Q^2_{\rm min}=1$ GeV$^2$.
In our calculations we use the next-to-leading order (NLO) ZEUS2002\_TR proton PDFs 
and QCDNUM~16.12~\cite{Botje:2010ay} for the scale evolution of the PDFs.
Furthermore, for simplicity, we neglect the contributions from the NLO Wilson coefficients which 
are known to be small.
Note that the uncertainties due to the extrapolation of the PDFs into the small-$x$ region and the scale uncertainties are much larger.

\begin{figure}[!h]
	\begin{center}
		\includegraphics[width=0.47\textwidth]{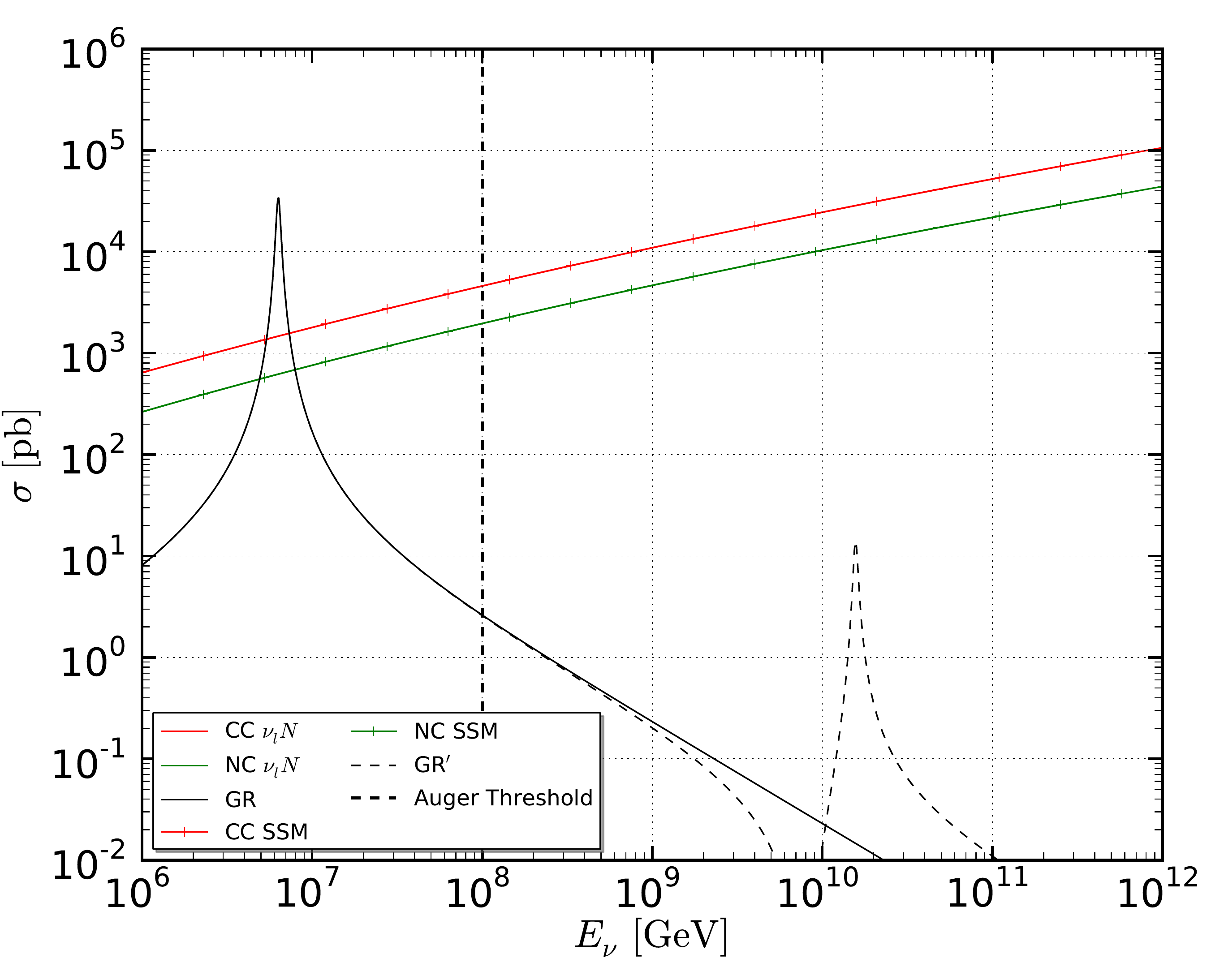}
	\end{center}\vspace{-0.5cm}
	\caption{Total cross sections for CC $\nu_\mu N$ DIS (red line), NC $\nu_\mu N$ DIS (green line) and the Glashow resonance (solid black line)
	in dependence of the incoming neutrino energy. The vertical line at $E_\nu = 10^8$ GeV indicates the lower energy threshold of the Auger Observatory.
The red and green crosses show the CC DIS and NC DIS cross sections, respectively, in the SSM with $M_{W'} = M_{Z'} = 4$ TeV.
The resonant $\bar\nu_e e^-$ scattering including the contribution from the $W'$ resonance is represented by the dashed, black line.}
\label{fig:cross-sections}
\end{figure} 

\begin{figure}[!h]
	\begin{center}
		\includegraphics[width=0.47\textwidth]{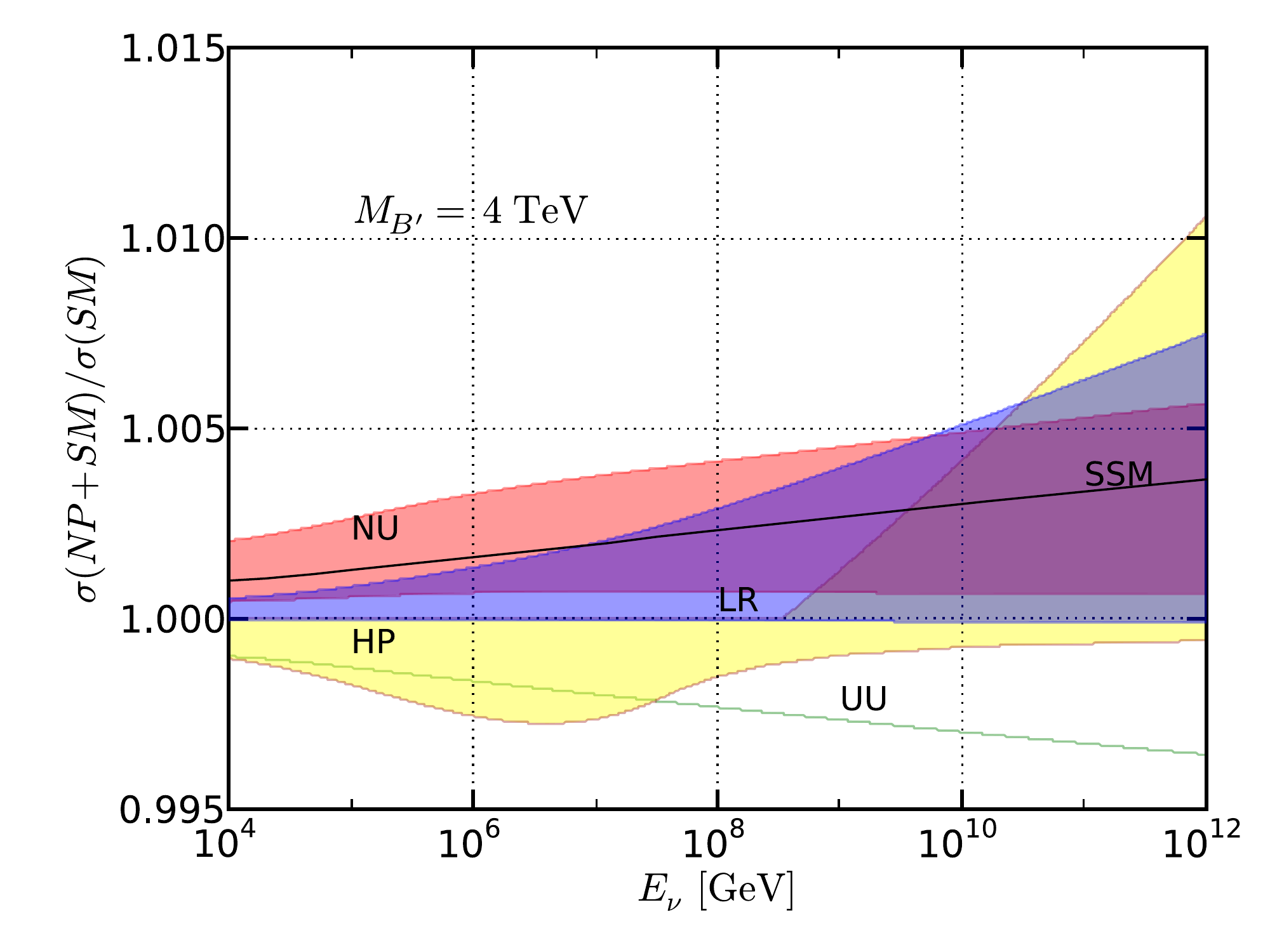}		
	\end{center}\vspace{-0.5cm}
	\caption{The CC+NC $\nu_\mu N$ DIS cross sections in different $\text{G}(221)$ models scaled to the cross section in the SM.
	The areas have been obtained by fixing either $M_{W'}=4$ TeV or $M_{Z'}=4$ TeV and scanning over the allowed parameter
	range of the model. For details we refer to Ref.\ \cite{Jezo:2012rm}. For comparison we also show the ratio obtained
	with the SSM using $M_{W'}=M_{Z'}=4$ TeV.}
\label{fig:ratio}
\end{figure} 

Our total cross sections for CC and NC DIS  are displayed in  Fig.~\ref{fig:cross-sections} as a function of the incoming neutrino energy $E_{\nu}$. 
We have verified that our cross section for CC DIS (red line) agrees with the results by Cooper-Sarkar et al. \cite{CooperSarkar:2011pa} 
within a few percent in the entire energy range shown.  It exceeds the CC cross section of Gandhi et al.  \cite{Gandhi:1998ri} 
by about 25\% at the highest energies $E_\nu=10^{12}$ GeV.
Conversely, our result for the NC cross section (green line) is 15\% - 20\% below the one in \cite{Gandhi:1998ri}.
In addition to the SM results, we present predictions for the total cross sections in the SSM (red and green crosses) 
assuming $M_{W'}=M_{Z'}=4$ TeV.
The DIS cross sections in the SM and the SSM differ at the 1\% level and the corresponding curves lie on top of 
each other. Similar observations hold for the other $\text{G}(221)$ models introduced above.
This can be seen in Fig.~\ref{fig:ratio}, where the ratio of the DIS cross sections in the new physics scenario and in the SM is presented.
The areas have been obtained by fixing, depending on the model, either $M_{W'}=4$ TeV or $M_{Z'}=4$ TeV and by
scanning over the allowed parameter spaces of the different models (see \cite{Jezo:2012rm} for details).
We find that the new physics contributions modify the SM results by at most 1\%, which is much smaller than the theoretical uncertainty
of the DIS cross sections. Similar results have been obtained for masses of the heavy resonance of 5 and 6 TeV.
We note that the ratio of the total cross sections could be enhanced by about ten percent
by imposing a suitable minimal $x_{\rm min}$-cut on the $x$-integration at the price of reducing the cross sections.
Indeed, the dominant contribution to the cross section comes from a region with ultra-small $x$-values 
(see Fig.~3 in \cite{Gluck:1998js}) and this region is shifted to larger $x$ due to the heavy resonance mass so that a cut on $x$ can considerably
reduce the SM DIS cross section while affecting less the result in the SSM.
For a similar reason, any suppression of the nuclear PDFs in the small $x$ region due to saturation effects would also lead to an enhanced signal 
to background ratio. However, an increase of the SM DIS cross section by 1 or 2\% is clearly not measurable with the Auger Observatory or any
foreseeable UHE neutrino experiment.

In Fig.~\ref{fig:cross-sections}, we also show numerical results for the production of hadrons
in resonant $\bar\nu_e e^-$ scattering in the SM (solid, black line) and in the SSM (dashed, black line). 
More specifically, we include the contributions with first and second generation quarks in the final state.
%
As can be seen, the GR cross section is more than one order of magnitude larger than the total CC neutrino DIS cross section 
at the resonance energy $E_\nu = 6.2 \cdot 10^6$ GeV. 
However, it decreases sharply away from the resonance, and the GR cross section is smaller than the 
CC DIS cross section by several orders of magnitude for energies greater than the Auger Observatory threshold, i.e. $E_{\nu} > 10^8$ GeV.
On the other hand, the contribution from the $W'$ resonance interferes destructively with the SM amplitude 
at energies below $10^{10}$ GeV but leads to a clear enhancement of the cross section in a 
bin around the $W'$-resonance energy $E_{\nu}^{\rm res}=M_{W'}^2/(2 m_e) \simeq 1.56 \cdot 10^{10}$ GeV.
Still it remains more than two orders of magnitude smaller than the DIS cross sections 
as can be inferred from Tab.~\ref{tab:x-sec} where we list the values of the different cross sections
at the peak of the resonance with mass $M_{W'}=4$ TeV.
For this reason, the effect of the GR${}^{\prime}$ resonance is irrelevant for events with hadronic showers.

\begin{table}[!h]
	\centering
	\scalebox{1.0}{%
	\begin{tabular}{|l|c|c|c|}
		\hline
	& Process & $\sigma$ [pb] (SM) & $\sigma$ [pb] (SSM)\\ \hline
1.) CC DIS & $\nu_\mu N \to \mu^- + X$ & $2.84 \cdot 10^4$  &  $2.84 \cdot 10^4$
\\
2.) NC DIS & $\nu_\mu N \to \nu_\mu + X$ & $1.20 \cdot 10^4$ & $1.20 \cdot 10^4$
\\
3.) GR${}^{(\prime)}$ to had.& $\bar\nu_e e^- \to {\rm hadrons}$ & $6.6 \cdot 10^{-2}$ & 41.16
\\
4.) GR${}^{(\prime)}$ to $e^-$& $\bar\nu_e e^- \to \bar\nu_e e^-$ & $1.1 \cdot 10^{-2}$ & 6.86
\\
5.) GR${}^{(\prime)}$ to $\mu^-$& $\bar\nu_e e^- \to \bar\nu_\mu \mu^-$ & $1.1 \cdot 10^{-2}$ & 6.86
\\
6.) ES into $e^-$ & $\nu_e e^- \to \nu_e e^-$, \ldots & 154.50 & ---
\\
7.) ES into $\mu^-$ & $\nu_\mu e^- \to \mu^-  \nu_e$ & 102.17 & ---
\\
\hline
	\end{tabular}
	}
	\caption{Cross sections at $E_\nu=1.56 \cdot 10^{10}$ GeV in the SM and the SSM assuming $M_{W'}=M_{Z'}=4$ TeV.
	The numbers in the 6th and 7th lines have been taken from figure 8 in \protect\cite{Gandhi:1998ri}.
	The elastic neutrino scattering off electrons into an electron (line 6) receives contributions from
	the following processes: $\nu_e e^- \to \nu_e e^-, \bar\nu_e e^- \to \bar\nu_e e^-, \nu_\mu e^- \to \nu_\mu e^-$,
	and $\bar\nu_\mu e^- \to \bar\nu_\mu e^-$. The non-resonant production of a muon (line 7) is due to the process 
	$\nu_\mu e^- \to \mu^- \nu_e$.
	}
	\vskip1em
	\label{tab:x-sec}
\end{table}

One way to enhance the relative importance of the new physics signal is to consider pure 'muon events' 
discussed in Ref.~\cite{Bhattacharya:2011qu} as a rather background free signal of the GR (in the SM).
The corresponding cross section for the resonant production of an electron or a muon is a factor 1/6 smaller than the
one shown in Fig.~\ref{fig:cross-sections} (see rows 3, 4, and 5 in Tab.~\ref{tab:x-sec}). 
As can be seen, at the resonance, the GR${}^{\prime}$ cross section in the SSM (row 5, column 4) is about 600 times larger than the one 
from the SM GR (row 5, column 3).
However,  it is necessary to take into account the non-resonant production of pure muon events
which, contrary to the SM case, is more important than the resonant mechanism.
The corresponding cross section in the SM, due to the process $\nu_\mu e^- \to \mu^- \nu_e$,
can be inferred from Fig.~8 in \protect\cite{Gandhi:1998ri}.
It depends only very mildly on the neutrino energy for $E_\nu > 10^8$ GeV and we provide its value at the
energy of the $W'$-resonance in row 7 of Tab.~\ref{tab:x-sec}.
For completeness, we also list the cross section for the elastic neutrino scattering in row 6.
We have not calculated the non-resonant elastic neutrino--electron scattering cross sections including additional
$W'$ and $Z'$ bosons but it is reasonable to assume that such contributions will modify the SM result at the low
percent level in the SSM and the $\text{G}(221)$ models when scanning over the allowed parameter range, similar
to the DIS case in Fig.~\ref{fig:ratio}. 
Therefore, we estimate that the contribution from the GR${}^{\prime}$ resonance enhances the cross section for muon
production in the SM by about 7\% at the resonance peak. 
Needless to say, that this enhancement gets reduced when calculating event numbers 
in appropriate energy bins.
In addition, we have estimated the background to the pure muon events due to CC DIS events where the hadronic shower energy is below 
the detection threshold which turns out to be much smaller than the signal so that it can be neglected.
However, the flux of UHE neutrinos will not be known with a better precision than 
the uncertainty of the DIS cross sections at very small $x$.
Therefore, it seems impossible for general reasons that the very precisely known leptonic cross sections
can be used to discover new spin-1 $W'$ and $Z'$ resonances.
In addition to these general considerations, the Auger Observatory has not yet detected UHE neutrino events.
A detector with a much larger acceptance would be required to measure the much smaller UHE neutrino--electron
cross sections. 

In conclusion,
we have computed UHE neutrino cross sections in the SSM and $\text{G}(221)$ models including additional
charged and neutral spin-1 resonances.
We find that the effects of such resonances are too small to be observed with the Auger Observatory or any
foreseeable upgrade of it. Conversely, should such resonances be observed at the LHC or a future hadron collider
they will have no measurable impact on the UHE neutrino events. Any deviation from the SM seen in
UHE cosmic neutrino events would require another explanation.

\acknowledgments
This work has been supported by a Ph.D.\ fellowship of the French
Ministry for Education and Research and by the Theory-LHC-France
initiative of the CNRS/IN2P3.

\bibliographystyle{apsrev4-1}
\bibliography{auger}
\end{document}